\documentclass[12pt,preprint]{aastex}
\usepackage{graphics,psfig,epsfig,lscape} 

\newcommand{\kms}{km\ s$^{-1}$}

\newcommand{\ii}{$i$}

\newcommand{\CO}{$^{12}$CO}
\newcommand{\CCO}{$^{13}$CO}
\newcommand{\COO}{C$^{18}$O}
\newcommand{\tx}{$\times$}

\slugcomment{ApJL, in prep.}
\shorttitle{AB Aurigae Resolved}
\shortauthors{Corder, Eisner \& Sargent} 

\begin{document}

\title{AB Aurigae Resolved: Evidence for Spiral Structure}
 
\author{Stuartt Corder, Josh Eisner, \& Anneila Sargent}
\affil{Division of Physics, Mathematics \& Astronomy, 
California Institute of Technology, MS105-24, Pasadena, CA 91125}
\email{sac,jae,afs@astro.caltech.edu}


\begin{abstract}

We obtained high angular resolution ($\sim$2$^{\prime\prime}$) images
of the \CCO(J=1$\rightarrow$0) line and 2.7 millimeter continuum
emission, and slightly lower resolution images of
\CO(J=1$\rightarrow$0) and \COO(J=1$\rightarrow$0) line emission
toward the Herbig Ae star AB Aurigae.  We resolve a circumstellar disk
of diameter 780 AU (FWHM) with a velocity pattern consistent with a
purely rotational disk at inclination 21.5$^\circ$ and position angle
58.6$^\circ$.  Using Keplerian disk models, we find a central source
dynamical mass of 2.8$\pm$0.1~M$_\odot$ and a cutoff radius of 615 AU
for the \CCO\ emission.  Inclination, mass, and radius determined from
\CO\ and \COO\ observations agree with those values, given optical
depth and abundance effects.  As a result of the high angular
resolution of our observations, we confirm the existence of spiral
structure suggested by near-IR scattered light images and show that
the spiral arms represent density contrasts in the disk.

\end{abstract}

\keywords{circumstellar matter --- stars: individual (AB Aur)}

\section{Introduction}\label{intro}

Circumstellar disks often surround low ($\lesssim$2~M$_\odot$) and
intermediate ($\sim$2-10~M$_\odot$) mass pre-main sequence objects,
the T Tauri (TT) and Herbig Ae/Be (HAeBe) stars.  Disk sizes are of
order a few hundred AU with masses a few tenths of a solar mass or
less (e.g. \citealt{beckwith96,natta00}). These values are similar to
the proto-solar system \citep{weidenschilling77}, suggesting that they
may be the sites of planet formation.  Disk temperature and density
profiles, key properties for understanding how planets might emerge,
have been inferred from spatially unresolved observations that rely on
spectral energy distributions (SEDs) and require assumptions about
disk morphology (e.g. \citealt{kenyon87,dullemond01}).  However,
higher angular and spectral resolution measurements of the gas and
dust in disks are critical to quantifying the profiles directly and
providing a context in which to interpret unresolved observations.
Spatially and kinematically resolved images enable measurement of
stellar mass, disk mass, radius, inclination (\ii), position angle
(PA), and substructure (e.g. \citealt{koerner93,dutrey98,simon01}).
Such observations of multiple spectral lines in DM Tau have allowed
the exploration of vertical and radial disk structure
\citep{dartios03}.

The HAe star AB Aurigae at 144 pc \citep{perryman97} has been studied
by numerous authors at optical, IR and radio wavelengths.  Most
recently, Semenov et al. (2004, hereafter S04) based a detailed
analysis on millimeter observations at 5-12'' resolution.  While this
angular resolution is insufficient to determine a stellar mass, their
model fits lead to a disk radius of 400$\pm$200 AU, \ii\ of
17$^{+6}_{-3}$$^\circ$, and PA of 80$\pm$30$^\circ$.  This inclination
agrees well with estimates of \ii$\lesssim$20-30$^\circ$ from near-IR
interferometry and scattered light
(\citealt{millan01,eisner03,eisner04,grady99,fukagawa04}, hereafter
F04).  Earlier millimeter interferometry and recent mid-IR
observations at angular resolution $\sim$0\farcs5 suggest
\ii$\sim$45-70$^\circ$ (\citealt{marsh95,mannings97b}, hereafter MS97;
\citealt{liu04}).

High resolution interferometric images of circumstellar gas can
determine the inclination definitively.  It is likely that early
studies of the AB Aur disk at moderate angular resolution were
affected by spiral structure in the disk surface (F04) or large scale
($\sim$1000 AU) envelope emission (e.g. S04).  Here we present high
resolution (2\farcs2) 2.7~mm continuum and \CCO(1-0) line images and
slightly lower resolution maps of the \CO(1-0) and \COO(1-0) emission
from AB Aur.  In addition to constraining the disk properties and
central mass, these enable us to detect and probe the spiral features
seen in the images of F04.

\section{Observations and Results}\label{obs_red}

Observations of AB Aur were carried out between January and May of
2004 at the Owens Valley Radio Observatory (OVRO) millimeter-wave
array.  Six 10.4 m antennas with cryogenically cooled SIS receivers
were used in five array configurations with baselines ranging from 15
to 480 m.  We configured the correlator to observe the emission lines
of \CCO\ at 110.2~GHz and \CO\ at 115.3~GHz in 64 Hanning-smoothed
channels of width 125~kHz, and the \COO\ emission line at 109.8 GHz in
32, 250~kHz wide channels.  The \CO\ and \COO\ observations have
maximum baselines of 103 and 120 m respectively.  Simultaneously, we
obtained 2.7~mm continuum measurements over 7 GHz of bandwidth with
OVRO's COBRA correlator \citep{hawkins04}.  The quasars J0530+135 and
3C111 were observed at 20 minute intervals for phase and amplitude
calibration.  Absolute flux calibration, accurate to 10\%, was based
on measurements of Uranus and Neptune.  Data reduction was carried out
with MIR, an IDL-based package developed for the OVRO array by
N. Scoville and J. Carpenter.  Maps were made using MIRIAD
\citep{sault95}.

\begin{figure*}
\includegraphics*[width=5.2 cm, angle=-0, clip=true]{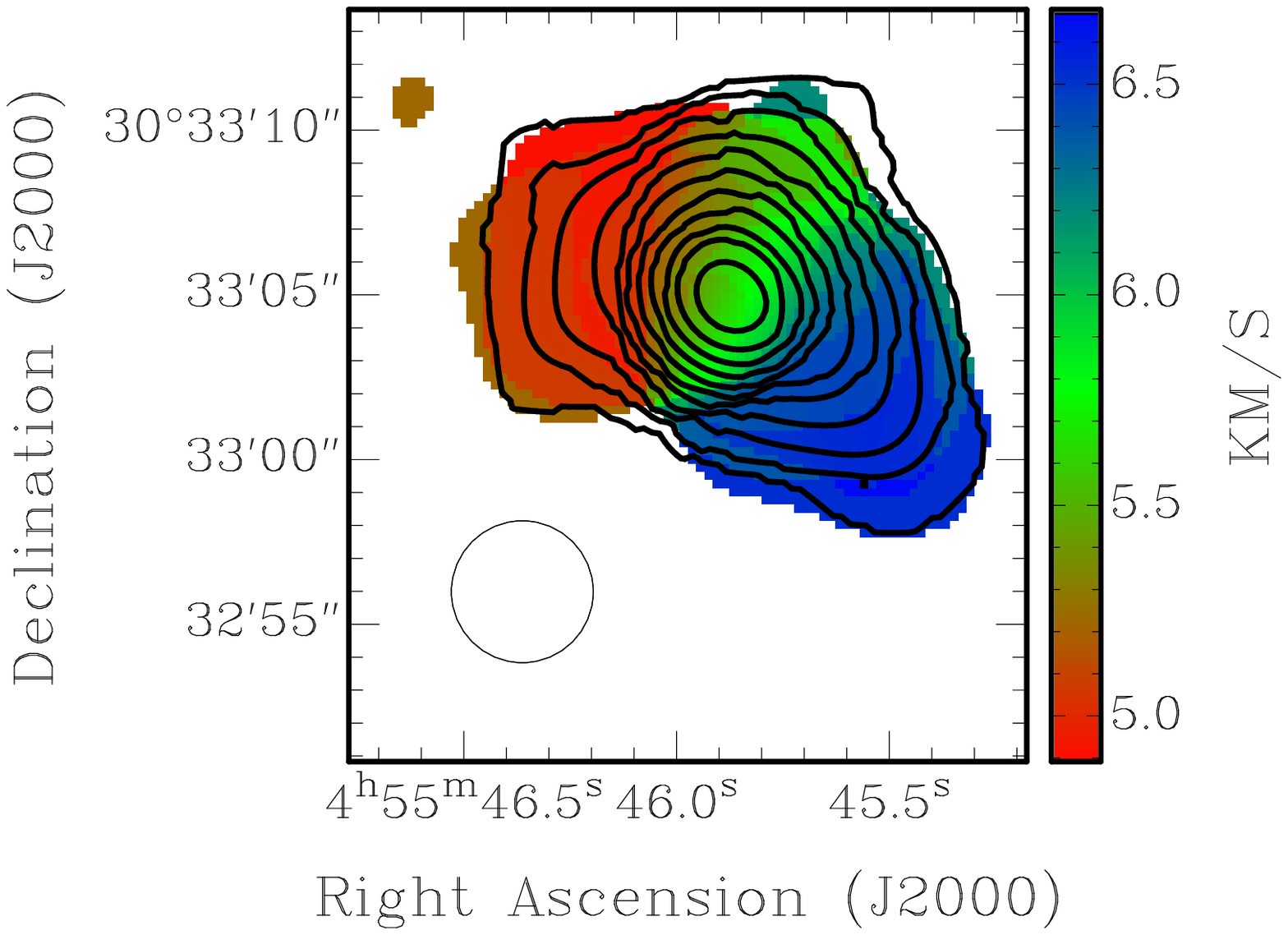}
\includegraphics*[width=5.2 cm, angle=-0, clip=true]{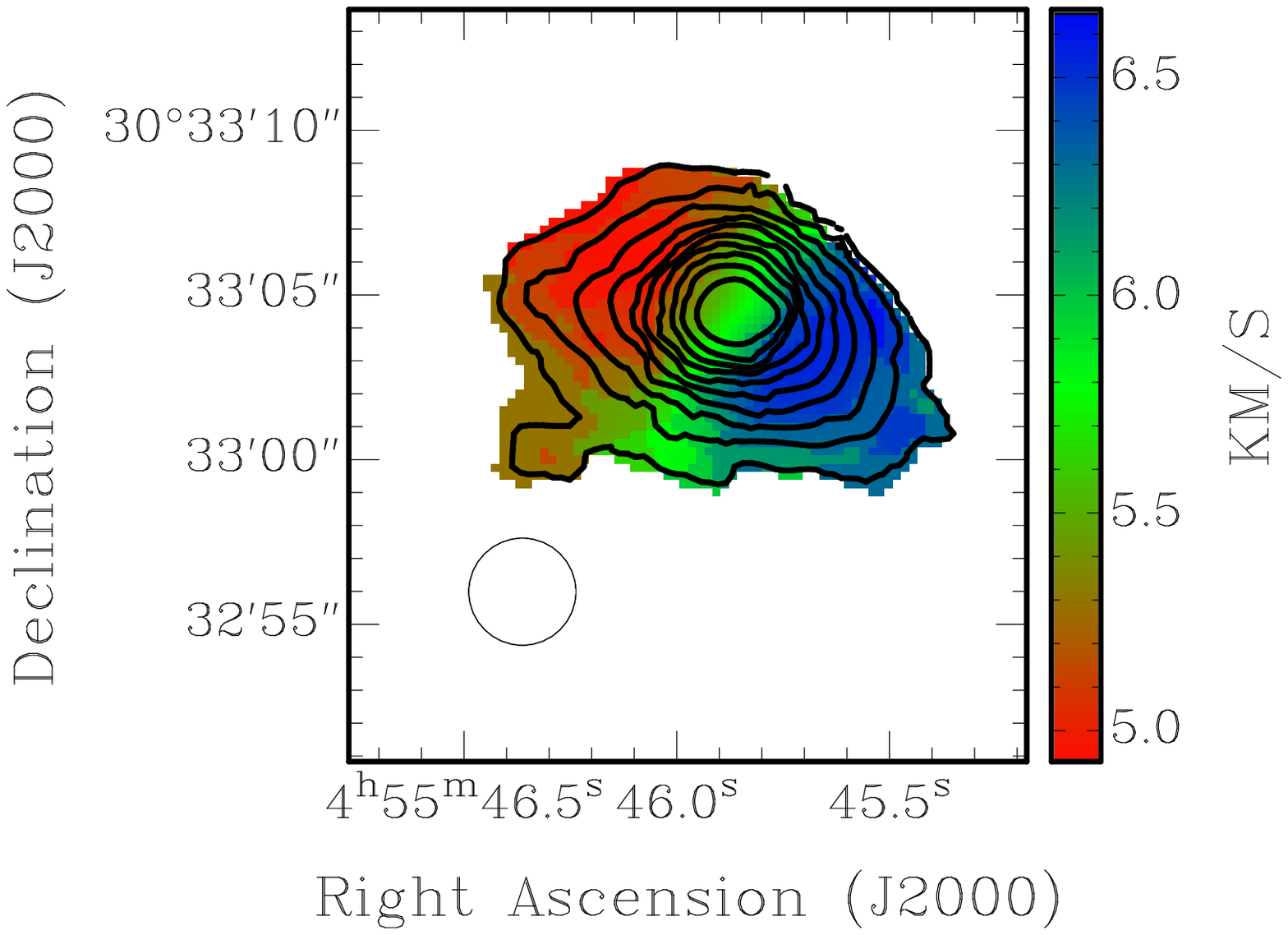}
\includegraphics*[width=5.2 cm, angle=-0, clip=true]{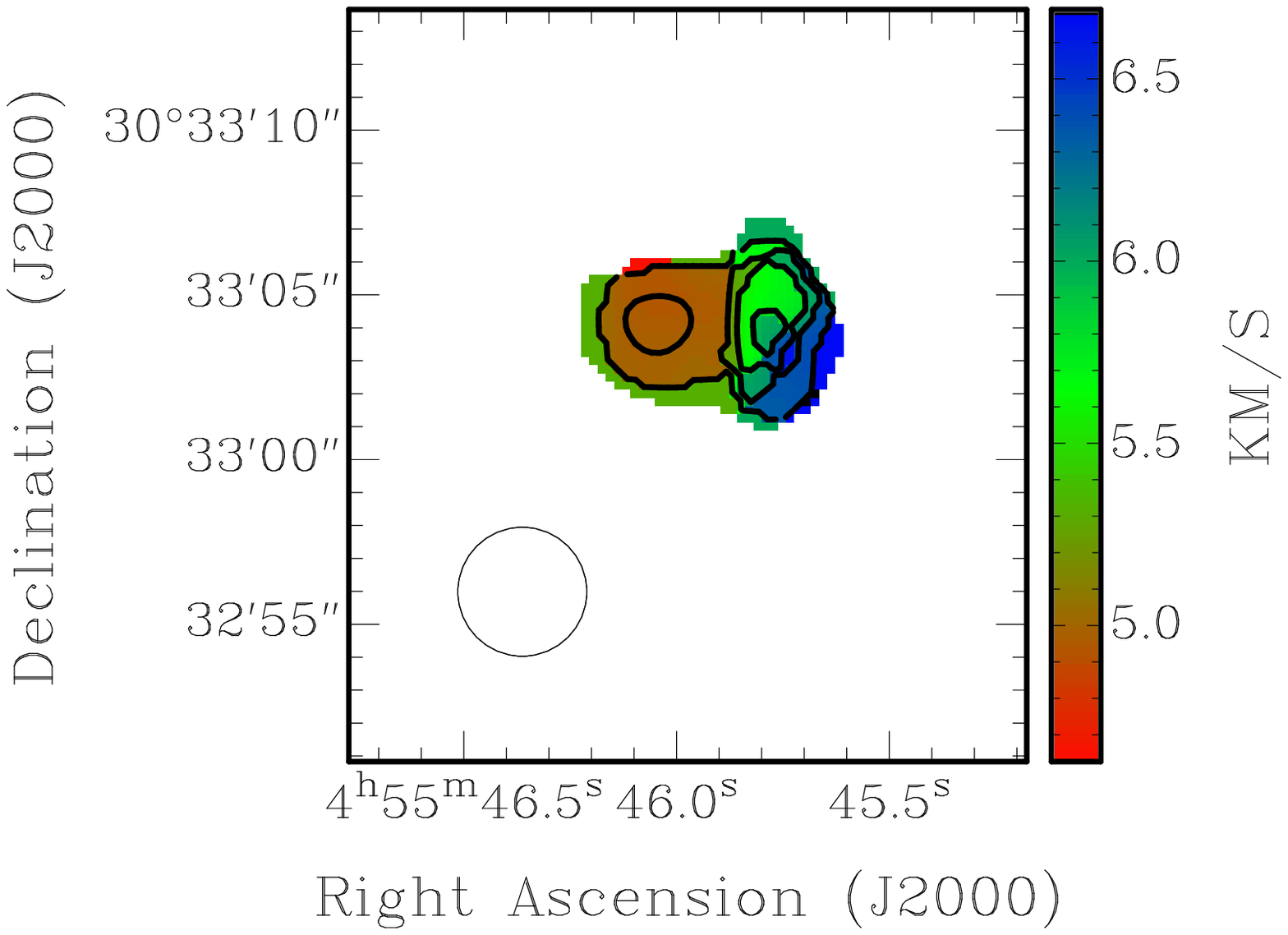}
\caption{OVRO millimeter array images of the \CO\ (left panel), \CCO\
(center panel) and \COO\ (right panel) emission from AB Aur.
Intensity-averaged velocity maps (first moment), shown in color, are
overlaid with contours of the integrated intensity (zeroth moment).
Contours begin at the 3$\sigma$ level and are spaced by 3$\sigma$ for
\CO\ and \CCO.  For \COO\, maps spacings are 2$\sigma$.  The
integrated \CO, \CCO, and \COO\ maps have 1$\sigma$ rms of 145, 36,
and 39~mJy~\kms\ respectively.  Emission was clipped at least at the
4$\sigma$ level in the first moment maps.  Circularly restored
naturally weighted equivalent beam sizes are shown in the bottom left
corners.
\label{moments}}
\end{figure*}

For AB Aur, Figure \ref{moments} shows contours of integrated
intensity for \CO~(left panel), \CCO~(center panel), and \COO~(right
panel) emission at resolutions 4\farcs25, 3\farcs25, and 3\farcs9,
respectively overlaid on intensity-weighted mean velocity maps
(color).  Integrated line fluxes for \CO, \CCO, and \COO\ are 22.1,
3.3 and 0.78~Jy~\kms\ over the velocity ranges 7.67 to 4.27, 7.51 to
4.25, and 7.37 to 4.63~\kms, respectively.  Outside these ranges, no
emission was detected above the 3$\sigma$ level (360, 105, and
72~mJy/beam for \CO, \CCO, and \COO\, respectively).  We combined our
\CCO\ dataset with that of MS97 (re-reduced using MIR) to provide
greater sensitivity and found the \CCO\ flux agrees with that of MS97,
within the uncertainties.  To emphasize global spatial and velocity
structure, the images were restored with circular beams equal in area
to the naturally weighted beams.  It is immediately evident that the
\CO\ and \CCO\ emission is resolved and the velocity varies smoothly
across the maps.  Due to lower sensitivity and poorer velocity
resolution, the \COO\ image is barely resolved and the velocity
pattern is sufficiently distorted to make analysis difficult.

In the left panel of Figure \ref{continuum_map}, contours of 2.7~mm
continuum emission at resolution 2\farcs2 are overlaid on the color
image of \CCO\ integrated intensity for AB Aur. Line and continuum
emission peaks are spatially coincident to within 0\farcs3.  The
optically thin nature of the 2.7~mm emission results in a direct
relationship of disk mass to flux, M$_d$=100$(F_{\nu}d^2/ \kappa_{\nu}
B_{\nu}(T))$.  Following MS97, we calculate the disk mass using this
expression and the 11.5~mJy 2.7~mm flux, assuming a gas to dust mass
ratio of 100 and $\kappa_\nu$=0.009~cm$^{2}$g$^{-1}$ at 2.7~mm.
Adopting a temperature of 40 K, we derive M$_d$=0.009~M$_\odot$ with a
factor of a few uncertainty.  S04 obtained a similar value,
M$_d$=0.013~M$_\odot$, with a factor of 7 uncertainty.  The center
panel of Figure \ref{continuum_map} displays our 2.7~mm continuum
contours overlaid on the near-IR scattered light image of F04, based
on which they suggest spiral structure.  While the scattered light
images trace small amounts of material, our 2.7~mm image demonstrates
that the spiral features are massive and represent density contrasts,
as the 2.7~mm flux directly traces the mass. The three most intense
scattered light features are detected at least at the 3$\sigma$ level
of significance in 2.7~mm continuum, while the strong northern and
southeastern scattered light arms are at better than the 5$\sigma$
level.  These northern and southeastern features are more obvious in
the right panel of Figure \ref{continuum_map}.  There, the symmetric
central core of the continuum image, containing more than 75\% of the
total flux, has been subtracted to enhance the outer regions.  Given
the uncertainties in the symmetric source fit, the resulting
northeastern and southeastern asymmetry residuals provide between 5\%
and 11\% and 7\% to 14\% of the total emission, respectively; residual
peak locations are uncertain by less than 0\farcs4, and emission peaks
are at least at the 3$\sigma$ level.

The smooth velocity gradient in the middle panel of Figure
\ref{moments} suggests that material is bound to the star and executes
Keplerian orbits, a hypothesis supported by S04.  The top panel of
Figure \ref{channel_maps} displays the morphology of the \CCO\
emission in velocity intervals of width 0.34 \kms.  The resolution of
these robustly weighted maps, 2\farcs2 \tx 1\farcs8, is almost a
factor of three better than achieved by MS97 and S04, as is the
signal-to-noise ratio.  Continuum resolution and signal-to-noise are
similarly improved.  Robust weighting allows us to emphasize small
scale features in individual velocity bins as opposed to global
characteristics.  Enhanced resolution and sensitivity enable a
detailed comparison of the line emission at different velocities with
that expected from models.  The model consists of a flat, Keplerian
disk with a single component power law emission profile and an outer
radius cutoff.  The emission profile was fit empirically and the
emission was binned by line-of-sight velocity and convolved with the
2\farcs2\tx 1\farcs8 beam.  The observed emission was subtracted from
the model and the goodness of fit was measured with a difference
squared merit function.  Our best fit model and 2$\sigma$ clipped
residuals are presented in the middle and bottom panels of Figure
\ref{channel_maps}.  Our interpretation relies entirely on the \CCO\
fit.  Given the poorer angular and spectral resolution of the \CO\ and
\COO\ observations, model fits for those lines were used as
consistency checks only.  The upper section of Table \ref{fit_result}
lists the best fit parameters for each spectral line.

The best fit dynamical mass for the \CCO\ data is
2.8$\pm$0.1~M$_\odot$.  Disk radius, inclination, and PA are
615$^{+8}_{-2}$ AU, 21.5$^{+0.4}_{-0.3}$$^\circ$ and
58.6$\pm$0.5$^\circ$ respectively.  The emission profile has form
$S=48.4(r/100)^{-1.28}$~mJy inside a radius of 615 AU.  Results for
the \CO\ and \COO\ are reasonably consistent.  We discuss any large
discrepancies below.  Estimates using other methods to obtain \ii, PA
and disk radius are shown in the lower section of Table
\ref{fit_result}.  The uncertainties quoted throughout this work are
purely statistical estimates, do not include possible errors in the
distance, and are 3$\sigma$ for our model fits and 1$\sigma$ for other
methods.  If the velocity pattern is not Keplerian the errors on all
disk parameters except radius could be substantially larger.  Disk
radius is more correlated with the assumed emission profile.

\section{Discussion}\label{discussion}

Our resolved images of the gas and dust emission from AB Aur enable us
to go beyond earlier millimeter observations at lower resolution and
determine a central mass of 2.8$\pm$0.1~M$_\odot$.  This dynamical
mass is only a little higher than the 2.4$\pm$0.2~M$_\odot$ obtained
from photometry and pre-main sequence track fitting \citep{ancker98}.
Our dynamical model results in several local $\chi^2$ minima but these
are significantly, statistically deviant ($\ge$20$\sigma$) from the
global minimum value.  Our best fit values of \ii, PA, and disk radius
(see Table \ref{fit_result}) are more accurate than those derived by
S04 from lower resolution data and more complex models, and well
within their uncertainties.

\begin{figure*}
\includegraphics*[angle=-90.0,clip=true,width=16.5 cm]{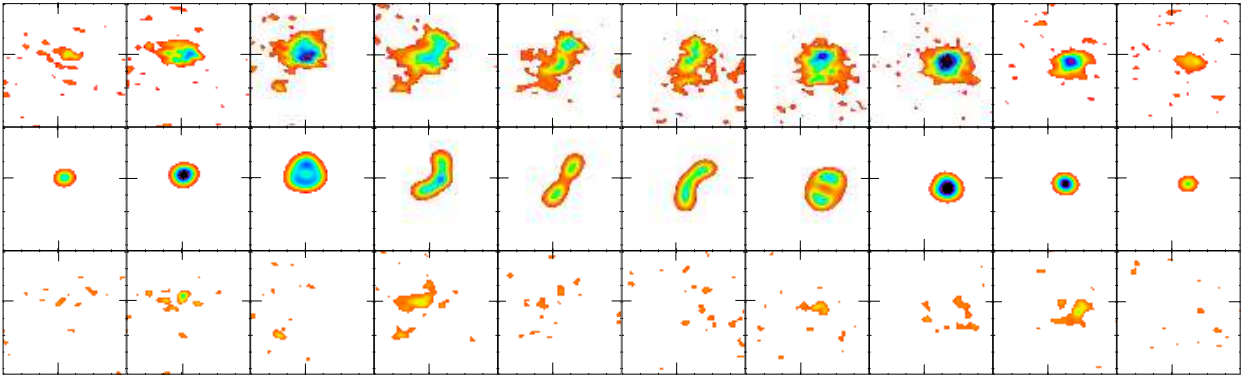}
\caption{Red, yellow, green, blue and dark blue emission represent
approximately 2, 3.5, 5.0, 6.5 and 8.5$\sigma$ above the noise.  The
left-most images have velocity 4.27~\kms.  Image velocity width is
0.34~\kms.  Large tick marks define the position
$\alpha=$04$^h$55$^m$46$^s$ and
$\delta=$30$^\circ$33$^\prime$05$^{\prime\prime}$.  Right ascension
ranges from 46.58$^s$ to 45.23$^s$ left to right and the declination
extends from 32$^\prime$54.8$^{\prime\prime}$ to
33$^\prime$12.3$^{\prime\prime}$ bottom to top.  Upper Panel: \CCO\
observations with 2\farcs2 by 1\farcs8 resolution and 1$\sigma$ rms
noise of 35~mJy/beam.  Middle Panel: Best fit Keplerian model with
mass$=$2.8~M$_\odot$, \ii$=$21.5$^\circ$, and PA$=$58.6$^\circ$.
Bottom Panel: Residuals of the data and model clipped at 2$\sigma$.
\label{channel_maps}}
\end{figure*}

Large scale envelope emission, present for r$>$600 AU (S04), may
contaminate more optically thick CO emission.  The best fit \CO\
emission profile described in Table \ref{fit_result} indicates
structure extending to beyond 1000~AU and flux ratios of CO
isotopomers show that \CO\ is optically thick.  Given this, the high
reduced $\chi^2$ value ($\sim$7) for our \CO\ model fit is not
surprising and demonstrates that a pure disk model does not fit the
\CO\ emission.  The largest deviation of \CCO\ from the model
($\sim$4.5$\sigma$) occurs in the second velocity bin (see Figure
\ref{channel_maps}).  The deviation is hardly significant and could be
due to a number of factors including local density enhancement, a
flared disk surface, etc.  A detailed discussion of these is beyond
the scope of this work, but there may be some evidence of disk flaring
in the top panel of Figure \ref{channel_maps} where the southern lobe
of emission in bins 4 through 7 splits into two pieces.  Given that
the \CCO\ observations fit the pure disk model much better than \CO\
and there is known contamination in the \CO\ emission, discrepancies
between disk properties derived from \CO\ and \CCO\ can be wholly
attributed to this effect, increasing our confidence in the \CCO\
values.

The model fits to mass and \ii\ based on \CCO\ and \COO\ emission are
consistent within the errors and the difference in radii is not
unexpected given the relatively low abundance of \COO.  Our PA
determinations are consistent with most published values.
Inclinations in the upper section of Table \ref{fit_result} are
consistent with near-IR interferometry results (Millan-Gabet et
al. 2001; \citealt{eisner03,eisner04}) for the inner 1~AU of the disk,
and with outer disk values from S04. We checked our model values by
calculating \ii\ using only the separation of the extreme velocity
emission and, as Table \ref{fit_result} shows, again find good
agreement. However, \ii\ values determined from axis ratios in the
visibility plane are systematically higher than the velocity based
methods.  Recent determinations of i$=$45-65$^\circ$ and
PA$=$30$\pm$15$^\circ$ by \citet{liu04} from mid-IR adaptive optics
observations with nulling interferometry, agree less well with our
model derived values, and their inclination is much closer to the
larger, axis-ratio-determined values of Table \ref{fit_result}.  We
believe that discrepancies between the values of inclination and
position angle derived by different methods are probably due to
assumptions about AB Aur symmetry.

\begin{figure*}
\includegraphics*[width=5.2 cm,angle=-0,clip=true]{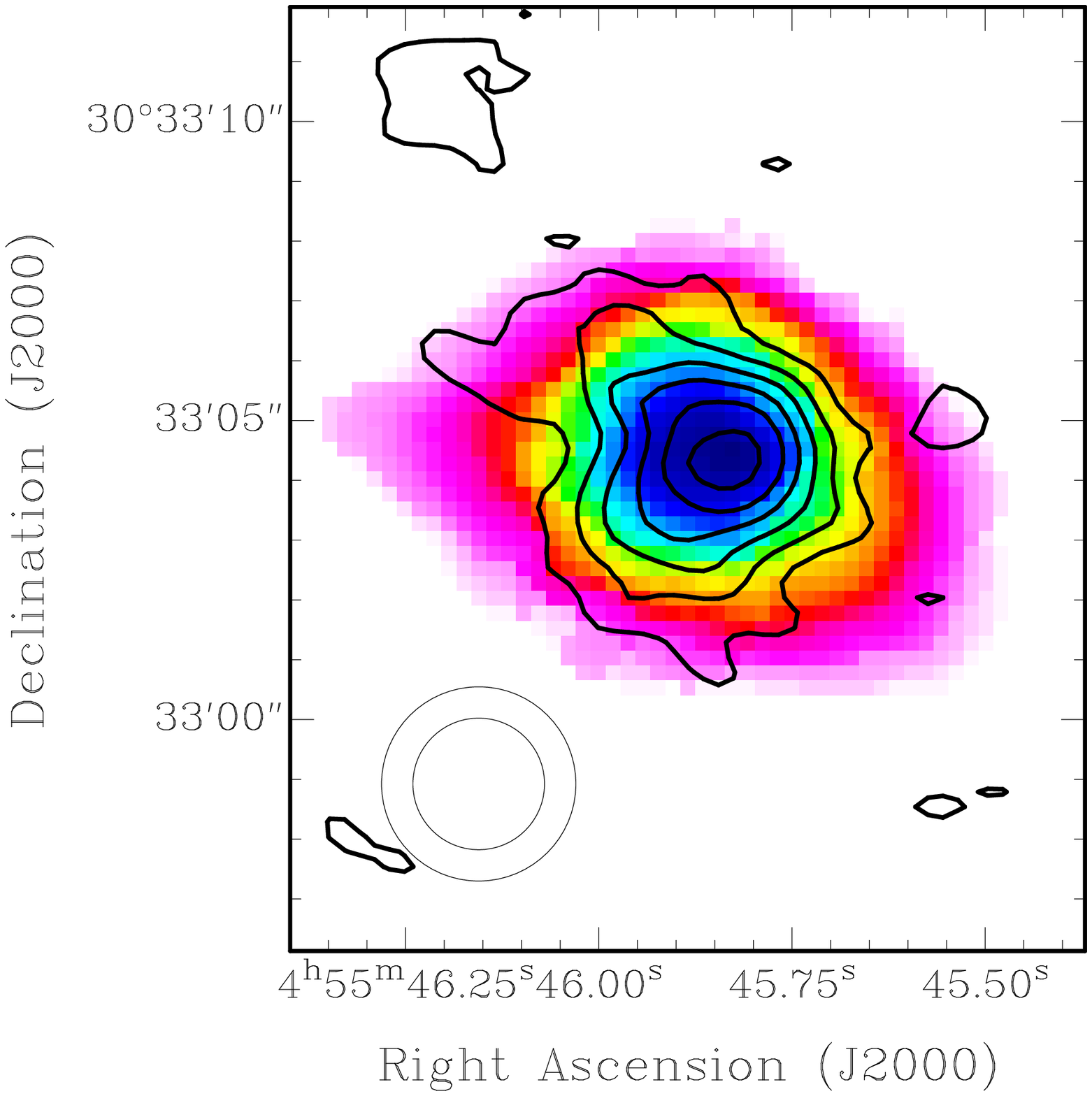}
\includegraphics*[width=5.2 cm,angle=-0,clip=true]{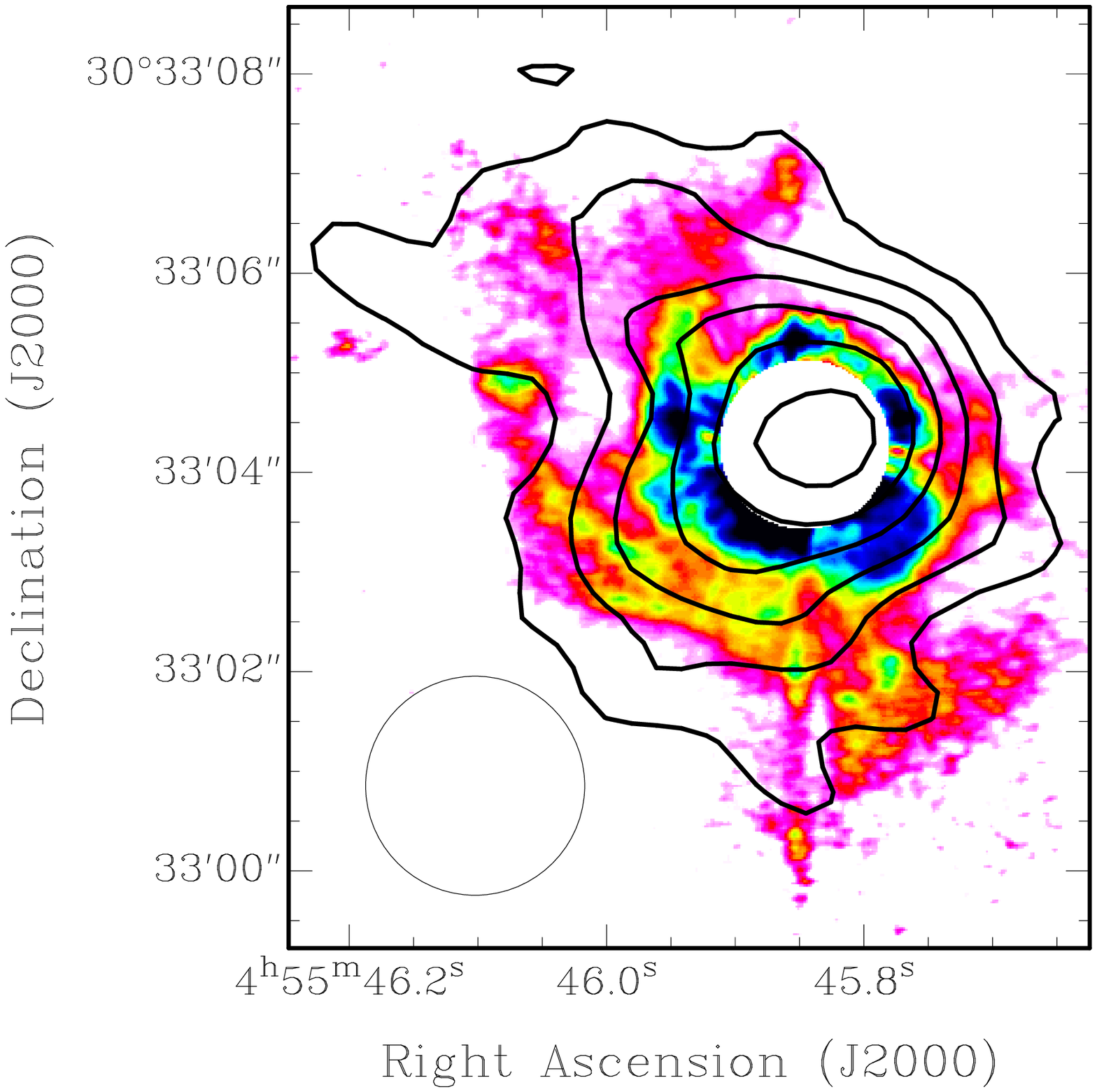} 
\includegraphics*[width=5.2 cm,angle=-0,clip=true]{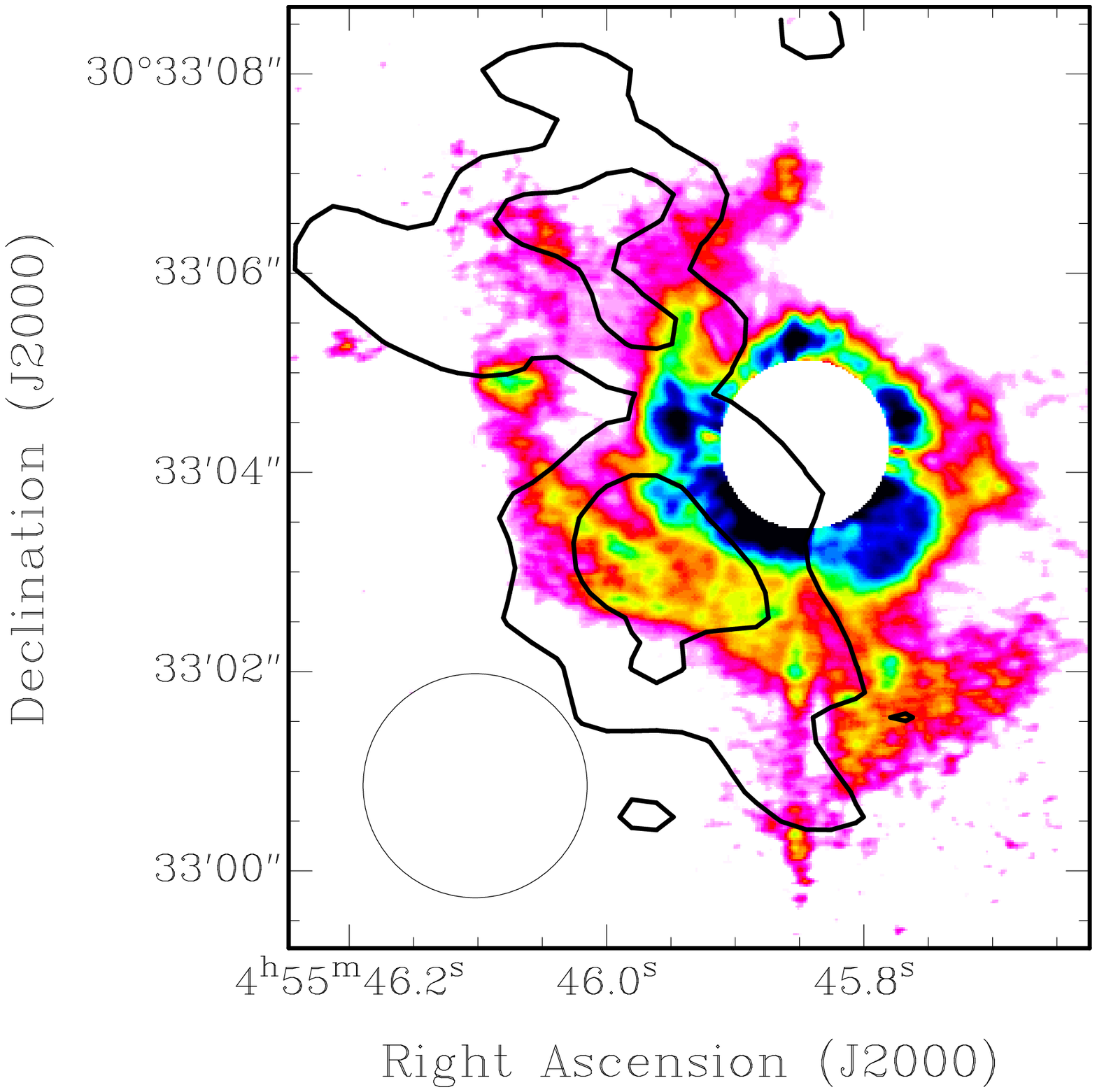}
\caption{Left panel: 2.7 mm continuum emission for AB Aur (contours)
overlaid on the \CCO\ integrated intensity image (color).  The lower
left portion of the map shows the continuum beam size.  The \CCO\ beam
surrounds the continuum beam in the left panel.  Visible \CCO\
emission begins at 3$\sigma$ level while the continuum contours begin
at 3$\sigma$ and increase as 2$\sigma$ where 1$\sigma$ is 0.36~mJy per
beam.  The center panel shows the same contours of this continuum
emission on the scattered light image of F04.  The right panel shows
contours of the continuum image with a best-fit gaussian source
subtracted.  Contours for this panel begin at 3$\sigma$ and increase
as 1$\sigma$.
\label{continuum_map}}
\end{figure*}

S04 and F04 demonstrate that the assumption of symmetry is invalid, as
do our millimeter observations.  The spiral structure, initially
suggested by F04 and demonstrated to be a density contrast through
this work (see Figure \ref{continuum_map}), is a major source of
asymmetry.  If spiral structure is present and streaming motions along
the arms are significant, the gas distribution could be tightened
along the arm, resulting in foreshortening perpendicular to that
direction.  Whether inclination increases or decreases depends on
source geometry.  In either case, the PA will be biased toward the
direction of the long axis of the spiral arm.  Spiral structure could
also affect the \ii\ and PA values determined from velocity structure,
depending on the strength of the features and the optical thickness of
the line; more optically thin emission features would experience
greater influence from such local density enhancements.  Sub-Keplerian
rotation will generally result in larger inclinations for velocity
determined methods with fixed mass, but a simple gas pressure aided
sub-Keplerian disk is probably too simplistic an assumption if
streaming motions are altering the bulk of the gas motion along the
arms.

Our observations show two effects that could be due to the influence
of spiral structure: 1) the model fit to the optically thin \COO\ line
leads to a significantly greater PA than does the \CCO\ fit.  The
direction of change of the PA is along the strong southeastern arm as
would be expected if local streaming velocities were playing a role;
2) in the left panel of Figure \ref{continuum_map}, the \CCO\ emission
spreads well beyond the boundaries of the continuum emission in the
northeast and southwest but has similar extent in the southeast.  If
spiral structure is changing gas morphology it would cause
\emph{larger} spatially determined inclinations and PAs thereby
explaining the higher inclinations estimated from axis ratios.  We
conclude that the range of published values of inclination and PA can
be explained by the presence of substructure.  The values derived from
our kinematic models may be influenced by this substructure, but we do
not expect the effects to be great.

The wealth of observations of AB Aur at different wavelengths and at
different spatial and spectral resolutions enables analysis of this
complicated circumstellar disk from the inner 1~AU (e.g.
\citealt{eisner04}) to the outer envelope beyond 1000 AU (e.g. S04).
Our detailed observations give an accurate dynamical mass and resolve
issues about \ii\ and PA, indicating that apparent inconsistencies are
due to asymmetry.  Our high resolution images in 2.7~mm continuum and
\CCO\ and \COO\ gas emission, show evidence of spiral structure.  Such
structure has important implications for planet formation. If it is
due to instability and persists long enough or is sufficiently strong,
local collapse can occur, quickly forming large planets
(\citealt{boss02,pickett03} and references therein).  Alternatively,
if the instability dissipates or the structure is due to existing
planets \citep{bate03}, the increase in local density may increase the
cross section for collisions of planetesimals and speed up the core
accretion process \citep{rice04}.

\acknowledgments Research with the OVRO array is funded in part by NSF
grant AST-9981546.  SC is supported by an NSF Graduate Research
Fellowship and JE by a Michelson Fellowship.  We thank M. Fukagawa for
providing her scattered light image of AB Aur, J. Carpenter for
interest and advice, and the referee, A. Natta, for constructive
comments.

\bibliographystyle{apj} \bibliography{resorbit}

\clearpage

\begin{large}
\end{large}

\clearpage



\begin{deluxetable}{crrrrrr}
\tabletypesize{\scriptsize} \tablecaption{Best-fit Model Parameters
\label{fit_result}} \tablewidth{0pt} \tablehead{ \colhead{Line} &
\colhead{Mass} & \colhead{Radius} &
\colhead{Inclination} & \colhead{Position Angle} & \colhead{Method} \\ 
\colhead{ \ } & \colhead{M$_\odot$} & \colhead{AU} & \colhead{Degrees} & \colhead{Degrees} & \colhead{ \ }} 
\startdata 

\CCO\ & 2.8$\pm$0.1 & 615$^{+8}_{-3}$ & 21.5$^{+0.4}_{-0.3}$ & 58.6$\pm$0.5 &  Model \\

\COO\ & 2.77$^{+0.09}_{-0.15}$ & 497$^{+10}_{-3}$ & 21.4$^{+0.7}_{-0.3}$ & 75$\pm$2 & Model \\ 

\CO\ & 3.25$^{+0.13}_{-0.14}$ & 1060$\pm$10 & 32.5$^{+0.4}_{-0.5}$ & 61$\pm$3 &  Model \\ 

\hline \hline \newline

\CCO & 2.79 & N/A & 24$\pm$4 & 69$\pm$8  & Extreme Velocity \\ 
\COO & 2.79 & N/A & 17$\pm$6 & 73$\pm$21 & Extreme Velocity \\ 
\CO  & 2.79 & N/A & 26$\pm$9 & 66$\pm$25 & Extreme Velocity \\ 

\CCO & N/A & 390$\pm$15 & 33$\pm$5  & 85$\pm$10 & UV Plane \\ 
\CO  & N/A & 500$\pm$13 & 41$\pm$3  & 26$\pm$4  & UV Plane \\ 
Dust & N/A & 170$\pm$17 & 44$\pm$12 & 79$\pm$19 & UV Plane \\

\enddata

\tablecomments{Dust is the 2.7~mm continuum emission.  Error bars are
1$\sigma$ for all but the line model where errors are 3$\sigma$.
Radii are HWHM values except for the line models where they represent
outer edges of the disk.  Emission profile for the line model was of
the form S=S$_0$(${r\over 100 AU}$)$^{-k}$~mJy for r$\le$R$_{out}$.
Values of (S$_0$,k) are (48.4$^{+1.3}_{-0.8}$,1.28$\pm$0.04),
(13.4$^{+1.2}_{-1.9}$,1.30$^{+0.06}_{-0.08}$), and
(150$\pm$2,1.1$\pm$0.1) for \CCO, \COO, and \CO\ respectively.}

\end{deluxetable}

\end{document}